\documentclass[conference]{IEEEtran}
\IEEEoverridecommandlockouts

\usepackage{cite}
\usepackage{amsmath,amssymb,amsfonts}
\usepackage{algorithmic}
\usepackage{graphicx}
\usepackage{textcomp}
\usepackage{xcolor}
\usepackage{comment} 
\usepackage{makecell} 
\def\BibTeX{{\rm B\kern-.05em{\sc i\kern-.025em b}\kern-.08em
    T\kern-.1667em\lower.7ex\hbox{E}\kern-.125emX}}
\begin{document}

\title{A Collaborative Rehabilitation-Exercise Serious Game for People with Stroke
and their Caregivers: \\A Pilot Study\\

\thanks{This work was supported by The Office of Vice Provost for Graduate Education at Stanford University.  979-8-3315-8342-2/26/\$31.00 ©2026 IEEE}
}

\author{\IEEEauthorblockN{ Elizabeth D. Vasquez}
\IEEEauthorblockA{\textit{Dept. of Mechanical Engineering} \\
\textit{Stanford University}\\
Stanford, USA \\
vasqueze@alumni.stanford.edu}
\and
\IEEEauthorblockN{ Jonathan Siskind}
\IEEEauthorblockA{\textit{Dept. of Mechanical Engineering} \\
\textit{Stanford University}\\
Stanford, USA \\
jsiskind@stanford.edu}
\and
\IEEEauthorblockN{ Marion S. Buckwalter}
\IEEEauthorblockA{\textit{Dept. of Neurology and Neurological Sciences} \\
\textit{Stanford University Medical Center}\\
Palo Alto, USA \\
marion.buckwalter@stanford.edu}
\and
\IEEEauthorblockN{ Maarten G. Lansberg}
\IEEEauthorblockA{\textit{Dept. of Neurology and Neurological Sciences} \\
\textit{Stanford University Medical Center}\\
Palo Alto, USA \\
lansberg@stanford.edu}
\and
\IEEEauthorblockN{ Sean Follmer}
\IEEEauthorblockA{\textit{Dept. of Mechanical Engineering} \\
\textit{Stanford University}\\
Stanford, USA \\
sfollmer@stanford.edu}
\and
\IEEEauthorblockN{ Allison M. Okamura}
\IEEEauthorblockA{\textit{Dept. of Mechanical Engineering} \\
\textit{Stanford University}\\
Stanford, USA \\
aokamura@stanford.edu}
}

\maketitle

\begin{abstract}
Motivation to perform movement therapy and caregiver burnout are major challenges to post-stroke life. Serious games have been shown to support therapeutic tasks in people with stroke, but there are few activities that simultaneously support informal caregiver health, which is also impacted post-stroke. Here, we present a collaborative, mutually beneficial, serious game designed to support therapy for persons with stroke and also exercise for their informal caregivers. One player performs rehabilitative wrist movements  -- useful to people with stroke -- and the other performs a seated march exercise -- useful to informal caregivers -- via pedals or a keyboard to control their avatar. We present a pilot study with 6 healthy dyads to evaluate how exercise-based input of one player, the Pseudo Caregiver (PCG), impacts motivation and emotional experience in both the PCG and Pseudo Person with Stroke (PPS). While not statistically significant, we find that PCGs Interest subscale scores trended higher when using a pedal (the exercised-based input) compared to a keyboard, regardless of game play mode. PPSs' positive affect scale scores and Competence subscale scores trended higher when their partner played collaboratively with a pedal compared to a keyboard. These trends encourage future work toward incorporating an exercise-based device, such as a pedal, to enhance the emotional and motivational experience of rehabilitative serious games for people with different movement ability levels.

\end{abstract}

\begin{IEEEkeywords}
Exergame, Rehabilitation, Stroke.
\end{IEEEkeywords}

\section{Introduction}
Motivation is essential to sustain the amount of therapeutic movement needed for recovery \cite{macleanCriticalReviewConcept2000}. Serious games -- games designed to achieve a goal other than pure entertainment -- can motivate therapy in people with stroke by enabling therapeutic movements in entertaining virtual environments \cite{chenHomebasedTechnologiesStroke2019}. Previous work has found that collaborative serious games, in particular, can promote engagement, performance, and social interaction \cite{nehrujeeIncreasingMotivationTrain2023, pereiraUseGameModes2021, maceBalancingPlayingField2017}, which foster motivation. 
 
Including social elements in serious games, such as family engagement, has been suggested to promote motivation \cite{alankusCustomizableGamesStroke2010}. However, in practice, additional healthy players do not directly benefit from their participation, and the games themselves often have lower physical challenge to accommodate the person with stroke \cite{maceBalancingPlayingField2017}. Studies have also shown that informal caregivers, who are often family members, face caregiving burdens that limit their available time, impact their health \cite{lohGlobalPrevalenceAnxiety2017}, create negative changes in their social environment \cite{andersonPopulationBasedAssessmentImpact1995}, and negatively impact their relationship with the person with stroke \cite{garciaCaregivingContextEthnicitya}. These burdens make it difficult to include family caregivers in movement therapy for people with stroke.
 
We propose supporting both the person with stroke and their informal caregiver through a collaborative serious game that allows the person with stroke to perform therapy while the informal caregiver performs a healthy behavior for themselves, exercise. This serious game draws on the social motivations of incorporating the family and aims to support more positive interactions and improved health for both players. To our knowledge, this is the first game that supports the health of both the person with stroke and their informal caregiver. Here, we present results from a pilot study with 6 dyads of healthy, acquainted volunteers to explore this system's impact on performance, positive affect, and motivation.

\section{Methods}
Although the eventual goal of this work is to determine the impact of caregiver exercise input in collaborative therapy games in people with stroke and their informal caregivers, this study first presents feasibility of this concept through healthy participants.

\subsection{Participants}
 Six dyads of healthy participants with no neurological conditions or injuries to their hands, arms, or feet participated in this study. Participants in each dyad -- a group of two people of close personal relation who completed the study together -- were recruited through convenience sampling and randomly assigned into either the Pseudo Person with Stroke (PPS) or Pseudo Caregiver (PCG) role and corresponding input device(s). A close personal relation was defined as family or a significant other of at least two years. During the study, participants were informed of their roles as either PPS or PCG. All participants were consented according to our approved Stanford Internal Review Board protocol (IRB-76985). See Table~\ref{tab:demographics} for self-described, demographic data. 

\begin{table}[tb]
\caption{Dyad Demographics}
\begin{tabular}{ccccccc}
\multicolumn{1}{c}{\textbf{ID}} & \multicolumn{1}{c}{\textbf{Age}} & \multicolumn{1}{c}{\textbf{Race}} & \multicolumn{1}{c}{\textbf{Sex}} & \thead{\textbf{Years} \\\textbf{Known}} & \multicolumn{1}{c}{\textbf{Relation}} & \thead{\textbf{IOS} \\\textbf{Change}}  \\ \hline
D0S0 & 28 & W** & M & 9 & P & 1 \\
D0C0 & 30 & W** & F & 9 & P & 0 \\
D1S1 & 25 & W & F & 5 & S & 0 \\
D1C1 & 25 & A & M & 5 & S & 0 \\
D2S2 & 25 & A & M & 2 & P & 0 \\
D2C2 & 25 & A & F & 2 & P & 0 \\ 
D3S3 & 22 & A & F & 3 & P & 1 \\
D3C3 & 24 & W & M & 3 & P & 0 \\
D4S4 & 27 & W* & F & 7 & S & 0 \\
D4C4 & 27 & MR* & M & 7 & S & 0 \\
D6S6 & 23 & W* & M & 3 & P & 0 \\
D6C6 & 22 & A, W, MR  & F & 3 & P & 0 \\ \hline          
\end{tabular}
\begin{center}
    \begin{scriptsize}
    Abbreviations: A = Asian, W = White, MR = More than one race, M = Male, \\ F = Female, P = Partner, S = Spouse, IOS = Inclusion of Other in the Self Scale, \\ Years Known = Self reported years the participant has known their partner, \\ IOS Change is Post minus Pre IOS (Positive IOS Change indicates an increase in the perception of closeness to their partner) * participant identified as Hispanic, \\ ** participant identified as Arab.
    \end{scriptsize}
    \end{center}
    \vspace{-0.4 cm}
\label{tab:demographics}
\end{table}

\subsection{Game Play Mechanics}
\begin{figure}[tb]
    \centering
    \includegraphics[height=4.2cm]{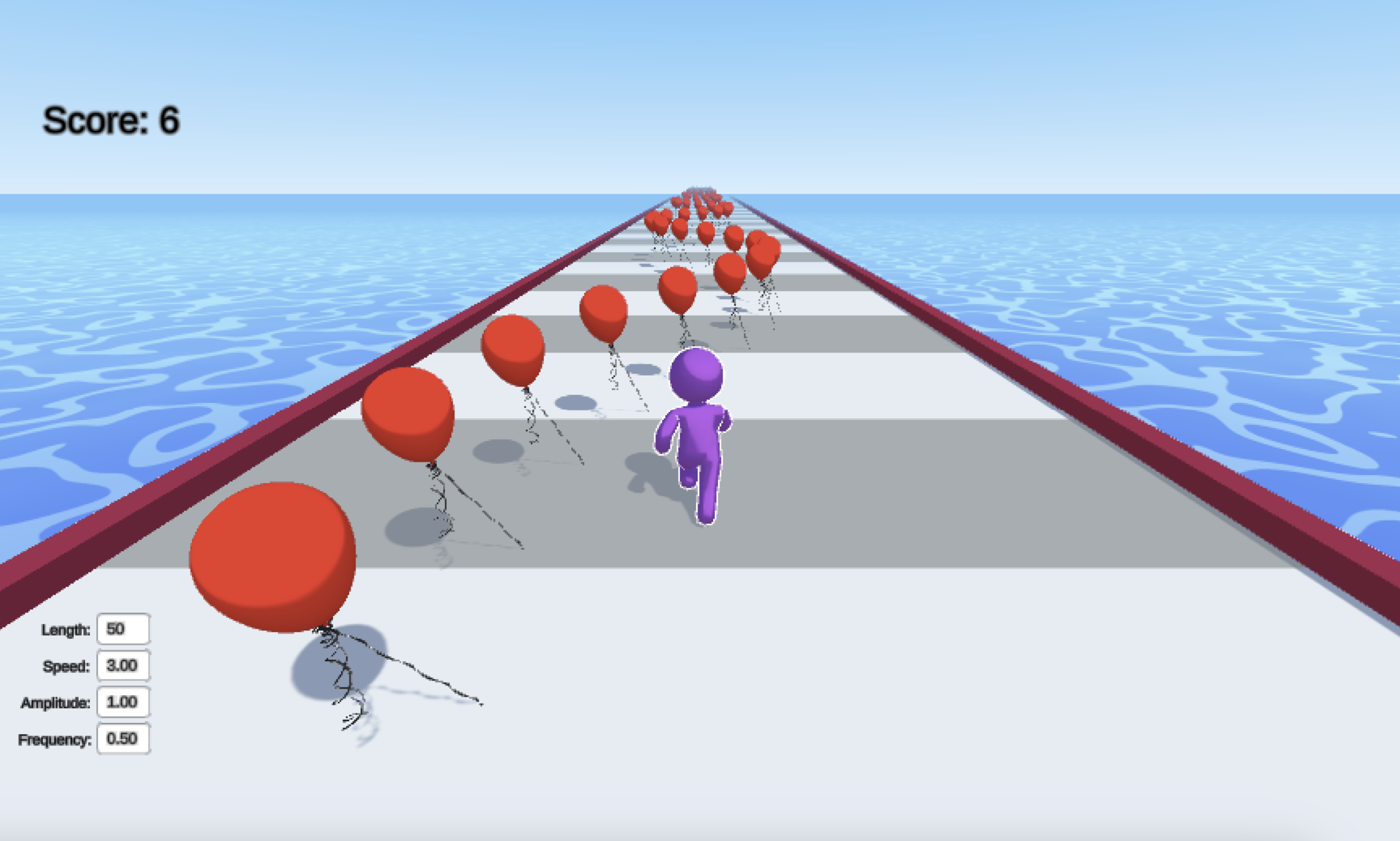} %
    \caption{\textbf{Solo Mode:} Solo player game mode is shown.}%
    \label{fig:solo}%
\end{figure}

\begin{figure}[t]
    \centering
    \includegraphics[height=4.2cm]{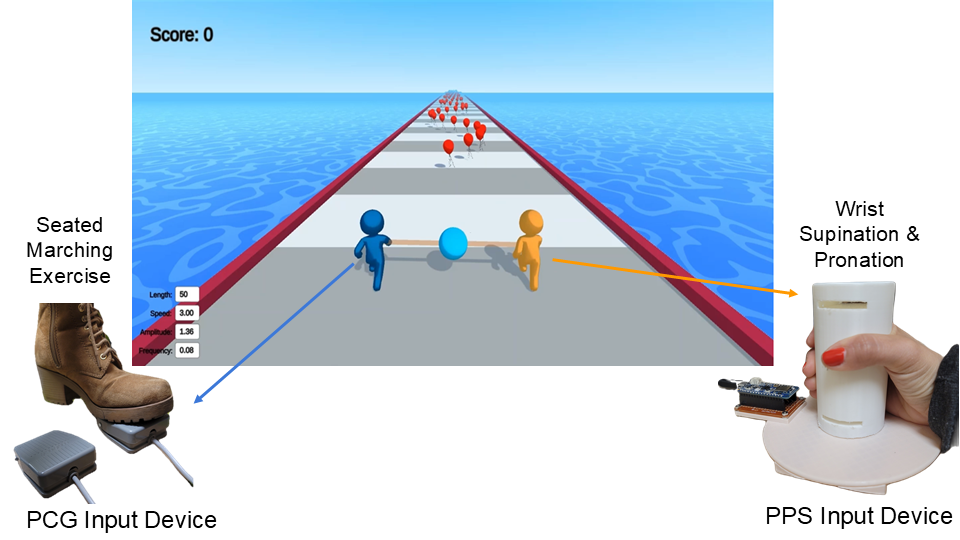} %
    \caption{\textbf{Game System:} Collaborative game mode is shown with the exercise input device used by the PCG on the left and the rehabilitative input device used by the PPS on the right.}%
    \vspace{-0.1 cm}
    \label{fig:game}%
\end{figure}

In the study, participants played an endless runner game similar to Nehrujee \textit{et al.} \cite{nehrujeeIncreasingMotivationTrain2023} with two game play modes. In Solo Mode, shown in Fig.~\ref{fig:solo}, participants play the game alone, and the objective is to collect balloons throughout the track by moving their game avatar to the balloons. In Collaborative Mode, shown in Fig.~\ref{fig:game}, the objective is the same, but the avatars of two players are connected by a line to show the distance between the players. Points are scored when the blue ball, which represents the average position of both players, contacts a balloon. If the avatars are less than one third of the track width apart from each other, the blue ball is large. If the avatars are between one third and two thirds of the track width apart from each other, the blue ball shrinks proportionally to the distance between the two avatars. If the avatars are more than two thirds of the track width apart from each other, the ball disappears and players can no longer collect balloons or score points until their avatars are less than two thirds of the track width apart – this encourages the players' avatars to stay together during game play. 

Balloons appear according to four randomized sinusoids of the form:
\[
\ x = A \sin(2\pi \cdot f \cdot z)
\]
where $x$ is the horizontal position, $A$ is the amplitude of the sine wave, $f$ is the frequency of the sinusoid (fixed at 0.08), and $z$ is the depth into the screen. The sine wave pattern extends into the $x-z$ axes in the game environment, and the avatar(s) run forward in the positive $z$ direction. In Solo Mode, the PCG's balloon amplitude, in meters, is randomly selected as: 0.75, 1.0, 1.25, or 1.5. The PPS's balloon amplitude is randomly selected as: 0.5, 0.7, 0.9, or 1.1. The PPS's possible amplitudes are scaled down 30 percent from the PCG's amplitudes, as the joystick has an approximately 30 percent reduced max range of motion compared to the uninhibited PCG's devices. The study proctor explained to participants that the slower movements of the PPS's avatar were designed to mimic the reduced mobility of a person with stroke. Each of the four possible balloon amplitudes appeared twice, once for practice and once for performance analysis. The maximum score possible across all modes and trials was 92, and a single point was scored for each balloon collected.

\subsection{PPS Rehabilitative Input Device}
The PPS input device consists of a 3D-printed joystick handle with an IMU (Adafruit Feather nRF52840 Sense) whose signal is passed through a Mahony Orientation Filter. By supinating and pronating their wrist, the user performs therapeutic wrist movements shown to support the therapeutic process of wrist recovery \cite{lambercyRehabilitationGraspingForearm2009}. This joystick-based device would best support people with stroke for whom wrist movements are achievable. When the handle grip is tilted to the left, the avatar translates left, and when the handle grip is tilted to the right, the avatar translates right. There is a deadzone in the center between -20 and 20 degrees where the avatar has zero lateral movement to simulate the reduced movement of a person with stroke. Past this threshold, noise was introduced to the lateral movement of the avatar by randomly altering its left/right speed to ensure that the player did not easily ``learn'' the system. To create this artificial noise, the player's lateral translation speed was randomly changed every 0.5 seconds to be between 0.5 and 1.5 $m/s$ while their forward speed remained constant at 3 $m/s$.

\subsection{PCG Exercise Input Device}
The exercise-based input device consists of two foot pedals, which the PCG used to move their character left or right. When using the pedals, PCGs were asked to perform a seated march exercise \cite{katoEffects12weekMarching2018} by tapping their left foot to the left pedal or their right foot to the right pedal. Holding down only the left pedal translates their avatar to the left at 1.5 m/s, and holding down only the right pedal translates their avatar to the right at 1.5 m/s. The avatar is always running forward at 3 m/s. Participants assigned to the PCG role also used keyboard arrows as our control variable for a non-exercise game input device.

\subsection{Experimental Procedure}
\begin{figure}[b]
    \centering
    \vspace{-0.4 cm}
    \includegraphics[width=1\columnwidth,trim={2pt 0 0 0},clip]{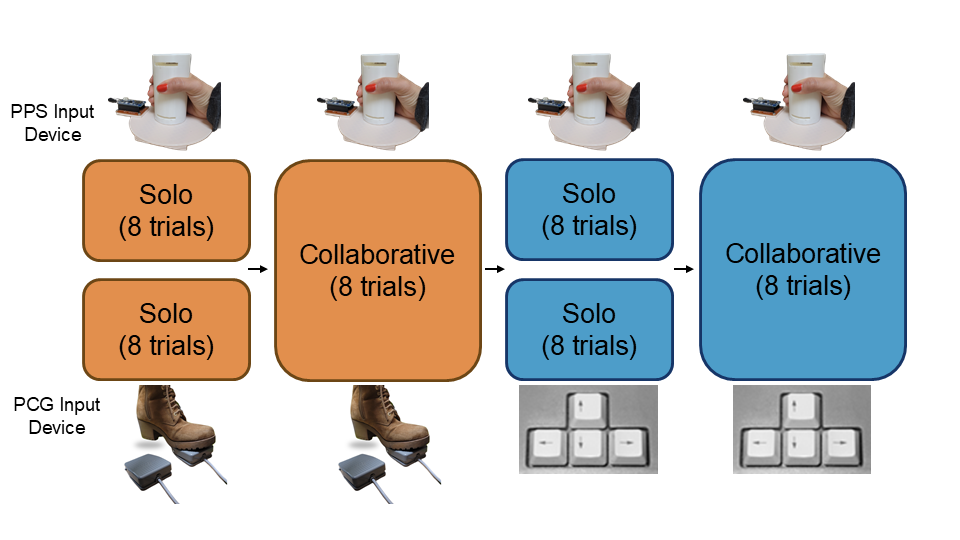}
    \caption{\textbf{Study protocol:} The study consists of four blocks. Each block is 8 trials, and each trial is 30 seconds long. The pedal and keyboard PCG input devices were counter balanced across participants.}
    \label{fig:protocol}%
\end{figure}

The study was composed of four blocks as shown in Fig.~\ref{fig:protocol}. Each participant played Solo Mode, Collaborative Mode, Solo Mode, and finally Collaborative Mode. In the first two blocks, the PCG input device was randomized as either the keyboard or pedal and the unselected input device was played for the last two blocks by the PCG. PPSs always played with the joystick input device. Each block had 8 trials, and each trial was 30 seconds long. The last four trials of each block were used to calculate performance. At the end of the study, participants were asked which game play condition -- mode and input device -- they preferred for themselves and which condition they thought their partner preferred. 

Before and after each block, participants completed the Intrinsic Motivation Inventory (IMI) \cite{marklandFactorialConstructValidity1997} and Positive and Negative Affect Schedule (PANAS) \cite{watsonDevelopmentValidationBrief1988}. The Inclusion of Others in Self Scale (IOS) \cite{aronInclusionOtherSelf1992} was administered at the start and end of the study as a closeness metric. 

The deviation of the participant's avatar to the balloon trajectory, Area Error, for each trial was calculated by numerical integration to find the absolute value of the area between the player's trajectory (in Solo Mode) or the ball's trajectory (in Collaborative Mode) and the sinusoidal path of the balloons. Trials with a lower Area Error indicate that the player(s) followed the balloon sinusoid more closely.

The Shapiro-Wilk Test was used to determine which test (t-test or Wilcoxon Test) to use for the following data: Performance (score and Area Error), IMI (Interest subscale, Competence subscale, and Tension subscale), and PANAS (positive affect scale and negative affect scale).

\section{Results}

\subsection{Performance}

\begin{table}[b]
    \caption{Performance Statistical Comparisons}
    \begin{tabular}{cccll}
        \thead{\textbf{Sample} \\\textbf{Population}} & \thead{\textbf{Groups} \\\textbf{Tested}} & \textbf{Metric} & \thead{\textbf{Test} \\\textbf{Statistic}} & \textbf{p-value} \\ \hline
        PCG & Mode & Score & t(5) = -5.85 & p $<$ 0.01* \\ 
        PCG & Mode & Area Error & z = -1.86 & p = 0.06 \\ 
        PCG (C) & Devices & Score & t(5) = -2.53 & p = 0.05* \\ 
        PCG (C) & Devices & Area Error & t(5) = 2.52 & p = 0.05* \\ 
        PCG (S) & Devices & Score & t(5) = -1.92 & p = 0.11 \\ 
        PCG (S)  & Devices & Area Error & t(5) = 4.04 & p = 0.01* \\ \hline
        PPS & Mode & Score & t(5) = -7.7 & p $<$ 0.01* \\ 
        PPS & Mode & Area Error & t(5) = 6.65 & p $<$ 0.01* \\ \hline
    \end{tabular}
    \begin{center}
    \begin{scriptsize}
    Abbreviations: (C) = Collaborative Mode, (S) = Solo Mode, \\  Mode = Game Play Mode (Solo, Collaborative), \\ Devices = PCG Input Devices (pedal, keyboard), \\ * indicates statistical or marginal significance, \\ Sample Population are the participants and experimental condition that both Groups Tested came from, Groups Tested are the groups the statistical test is performed between, Metric is the particular subscale score of interest.
    \end{scriptsize}
    \end{center}
    \vspace{-0.4 cm}
    \label{tab:performance}
\end{table}

In Solo Mode, PCGs, on average (Score: $83.63 \pm 8.84$, Area Error: $21.52 \pm 5.36$), scored higher than PPSs (Score: $78.15 \pm 11.82$, Area Error: $21.34 \pm 7.08$), which suggests that PPSs had more difficult game play than PCGs. Looking at performance differences by game play mode, both PCGs and PPSs scored significantly lower and had at least marginally significantly higher Area Error when playing in Collaborative Mode than Solo Mode regardless of the PCGs device (See Table~\ref{tab:performance}). Results demonstrate that when playing in Collaborative Mode, PCGs had marginally significantly lower scores ($t(5) = -2.53$, $p = 0.05$) and higher Area Error ($t(5) = 2.52$, $p = 0.05$) when playing with the pedal than with the keyboard. However, when playing in Solo Mode, PCGs' scores were not significantly lower playing with the pedal than keyboard, but their Area Error was significantly higher playing with the pedal than keyboard ($t(5) = 4.04$, $p = 0.01$). This suggests that the pedal was a more difficult input device to use than the keyboard and Collaborative Mode was more difficult than Solo Mode with the Mode difference having the stronger impact on performance.

\subsection{Positive and Negative Affect Schedule}

\begin{table}[tb]
    \caption{Positive and Negative Affect Schedue Statistical Comparisons}
    \begin{tabular}{cccll}
         \thead{\textbf{Sample} \\\textbf{Population}} & \thead{\textbf{Groups} \\\textbf{Tested}} & \textbf{Metric} & \thead{\textbf{Test} \\\textbf{Statistic}} & \textbf{p-value} \\ \hline
        PCG (S) & Devices & Positive & t(5) = 0.12 & p = 0.91 \\ 
        PCG (C) & Devices & Positive & t(5) = -0.40 & p = 0.70 \\ 
        PCG (S) & Devices & Negative & t(5) = -1.02 & p = 0.36 \\ 
        PCG (C) & Devices & Negative & t(5) = 0.30 & p = 0.77 \\ \hline 
        PPS (S) & Devices & Positive & t(5) = -0.35 & p = 0.74 \\ 
        PPS (C) & Devices & Positive & t(5) = 1.63 & p = 0.18 \\ 
        PPS (S) & Devices & Negative & z = -0.18 & p = 0.85 \\ 
        PPS (C) & Devices & Negative & t(5) = 1.20 & p = 0.29 \\ \hline
    \end{tabular}
    \begin{center}
    \begin{scriptsize}
    Abbreviations: (C) = Collaborative Mode, (S) = Solo Mode, \\ Positive = Positive Affect, Negative = Negative Affect \\ Devices = PCG Input Devices (pedal, keyboard) \\ Sample Population are the participants and experimental condition that both Groups Tested came from, Groups Tested are the groups the statistical test is performed between, Metric is the particular subscale score of interest.
    \end{scriptsize}
    \end{center}
    \vspace{-0.4 cm}
    \label{tab:panas}
\end{table}

\begin{figure}[b]
    \centering
    \includegraphics[width=1\columnwidth,trim={2pt 0 0 0},clip]{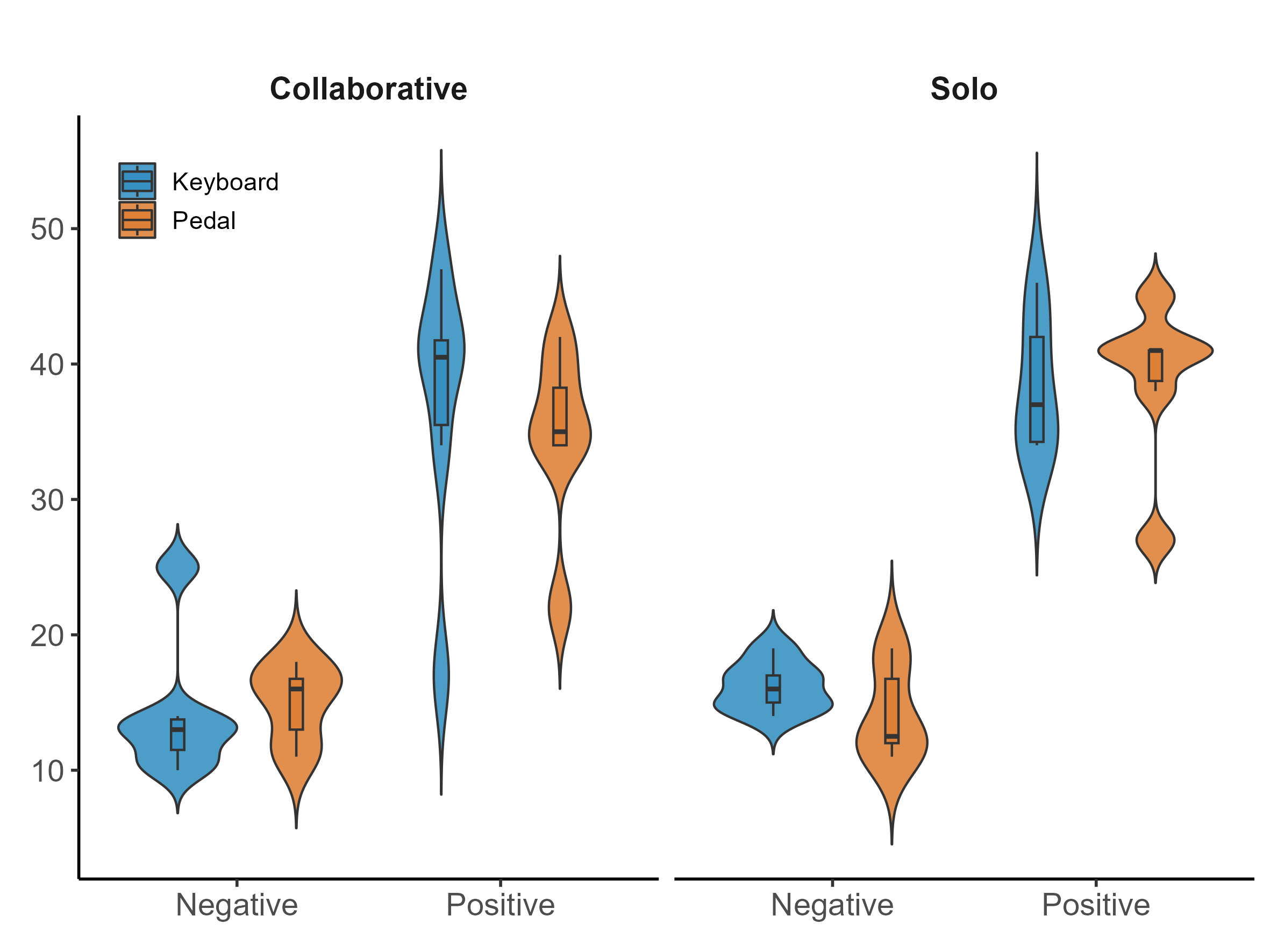}
    \caption{\textbf{PCG Positive and Negative Affect Schedule:} PCG scores trended toward more positive emotions and fewer negative emotions in the pedal than keyboard input device when playing Solo Mode. This was reversed in Collaborative Mode.}
    \label{fig:panas_cg}%
\end{figure}

 As shown in Fig.~\ref{fig:panas_cg}, the median PCG PANAS positive affect scale score trended, but was not significantly, higher when using the pedal ($41.0 \pm 6.2$) than the keyboard ($37.0 \pm 5.1$) in Solo Mode and trended lower when using the pedal ($35.0 \pm 6.9$) than the keyboard ($40.5 \pm 10.6$) in Collaborative Mode. The median PCG PANAS negative affect scale score trended lower when using the pedal ($12.5 \pm 3.4$) than the keyboard ($16.0 \pm 1.8$) in Solo Mode and trended higher when using the pedal ($16.0 \pm 2.8$) than the keyboard ($13.0 \pm 5.4$) in Collaborative Mode. As shown in Table~\ref{tab:panas}, there were no significant differences in PANAS positive or negative affect scale scores across input devices for either Solo or Collaborative Mode. 

\begin{figure}[tb]
    \centering
    \includegraphics[width=1\columnwidth,trim={2pt 0 0 0},clip]{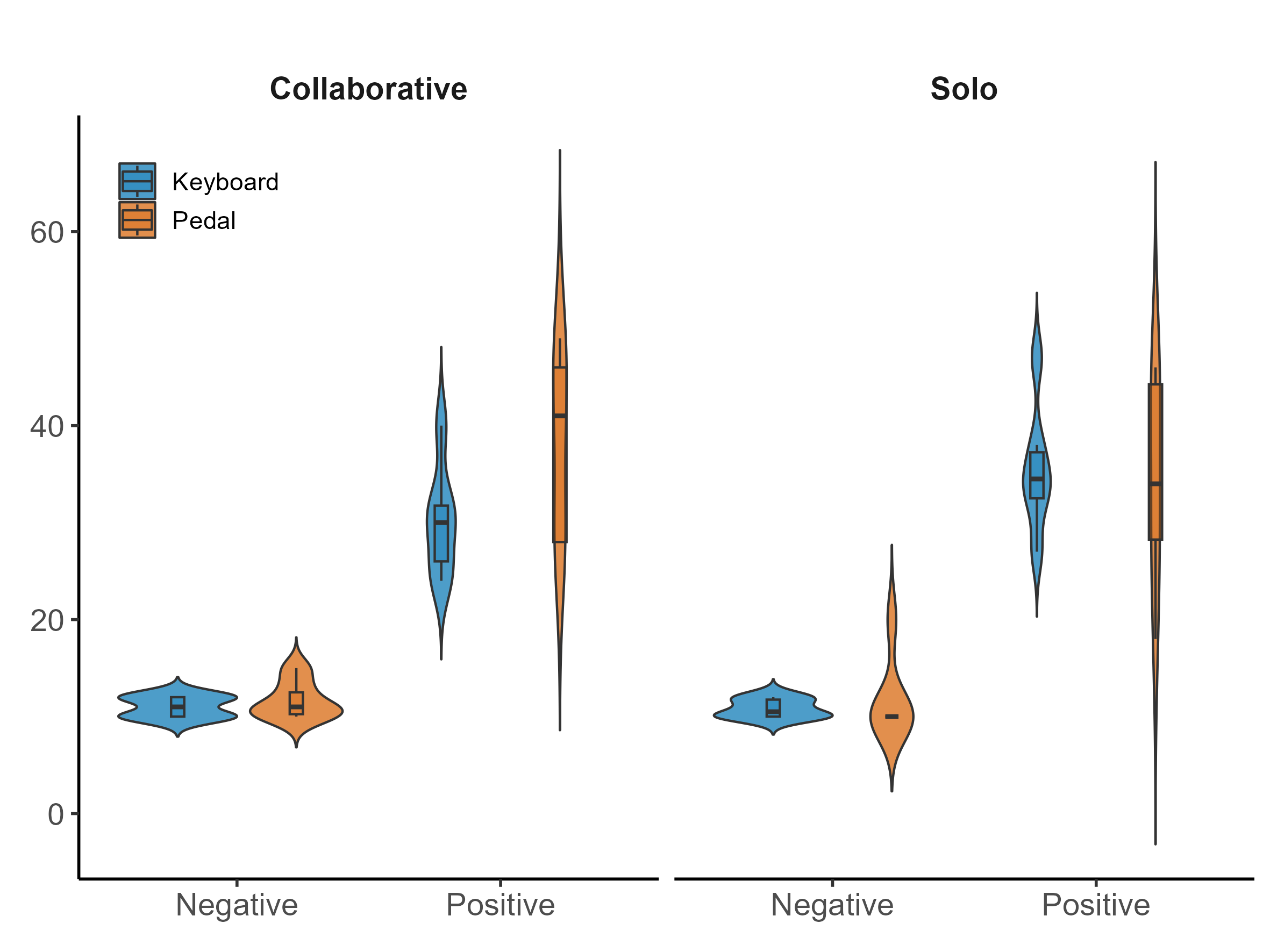}
    \caption{\textbf{PPS Positive and Negative Affect Schedule:} PPSs scores trended toward more positive emotions when PCGs played with the pedal in both play conditions.}
    \label{fig:panas_ss}%
\end{figure}

 As shown in Fig.~\ref{fig:panas_ss}, the median PPS PANAS positive affect scale score trended lower when the PCG was playing with the pedal ($34.0 \pm 11.2$) than the keyboard ($34.5 \pm 6.7$) in Solo Mode and trended higher when the PCG was playing with the pedal ($41.0 \pm 9.9$) than the keyboard ($30.0 \pm 5.8$) in Collaborative Mode. The median PPS negative affect scale score trended lower when the PCG used the pedal ($10.0 \pm 4.1$) than the keyboard ($10.5 \pm 1.0$) in Solo Mode and was the same (pedal: $ 11.0 \pm 2.0$, keyboard: $11.0 \pm 1.1$) in the Collaborative Mode. As shown in Table~\ref{tab:panas}, there were no significant differences in PANAS positive or negative affect scale scores across input devices for either Solo or Collaborative Mode.

\subsection{Intrinsic Motivation Inventory}

\begin{table}[!ht]
    \caption{Intrinsic Motivation Inventory Statistical Comparisons}
    \begin{tabular}{cccll}
         \thead{\textbf{Sample} \\\textbf{Population}} & \thead{\textbf{Groups} \\\textbf{Tested}}  & \textbf{Metric} & \thead{\textbf{Test} \\\textbf{Statistic}} & \textbf{p-value} \\ \hline
        PCG (S) & Devices & Tension & t(5) = -0.89 & p = 0.42 \\ 
        PCG (C) & Devices & Tension & t(5) = 1.73 & p = 0.14 \\ 
        PCG (S) & Devices & Interest & t(5) = 0.13 & p = 0.90 \\ 
        PCG (C) & Devices & Interest & t(5) = 0.50 & p = 0.64 \\ 
        PCG (S) & Devices & Competence & t(5) = 1.24 & p = 0.27 \\ 
        PCG (C) & Devices & Competence & z = -2.10 & p = 0.04* \\ 
        PCG (P) & Mode & Competence & t(5) = -2.57 & p = 0.05* \\ \hline
        PPS (S) & Devices & Tension & t(5) = 3.37 & p = 0.02* \\ 
        PPS (C) & Devices & Tension & t(5) = 0.47 & p = 0.66 \\ 
        PPS (S) & Devices & Interest & t(5) = -0.04 & p = 0.97 \\ 
        PPS (C) & Devices & Interest & t(5) = 0.53 & p = 0.62 \\ 
        PPS (S) & Devices & Competence & t(5) = -0.16 & p = 0.88 \\ 
        PPS (C) & Devices & Competence & t(5) = 1.27 & p = 0.26 \\ \hline
    \end{tabular}
    \begin{center}
    \begin{scriptsize}
    Abbreviations: (C) = Collaborative Mode, (S) = Solo Mode, \\ (P) = pedal condition only, Mode = Game Play Mode (Solo, Collaborative), \\ Devices = PCG Input Devices (pedal, keyboard), \\ * indicates statistical or marginal significance, \\ Sample Population are the participants and experimental condition that both Groups Tested came from, Groups Tested are the groups the statistical test is performed between, Metric is the particular subscale score of interest.
    \end{scriptsize}
    \end{center}
    \vspace{-0.4 cm}
    \label{tab:imi}
\end{table}

\begin{figure}[tb]
    \centering
    \includegraphics[width=1\columnwidth,trim={2pt 0 0 0},clip]{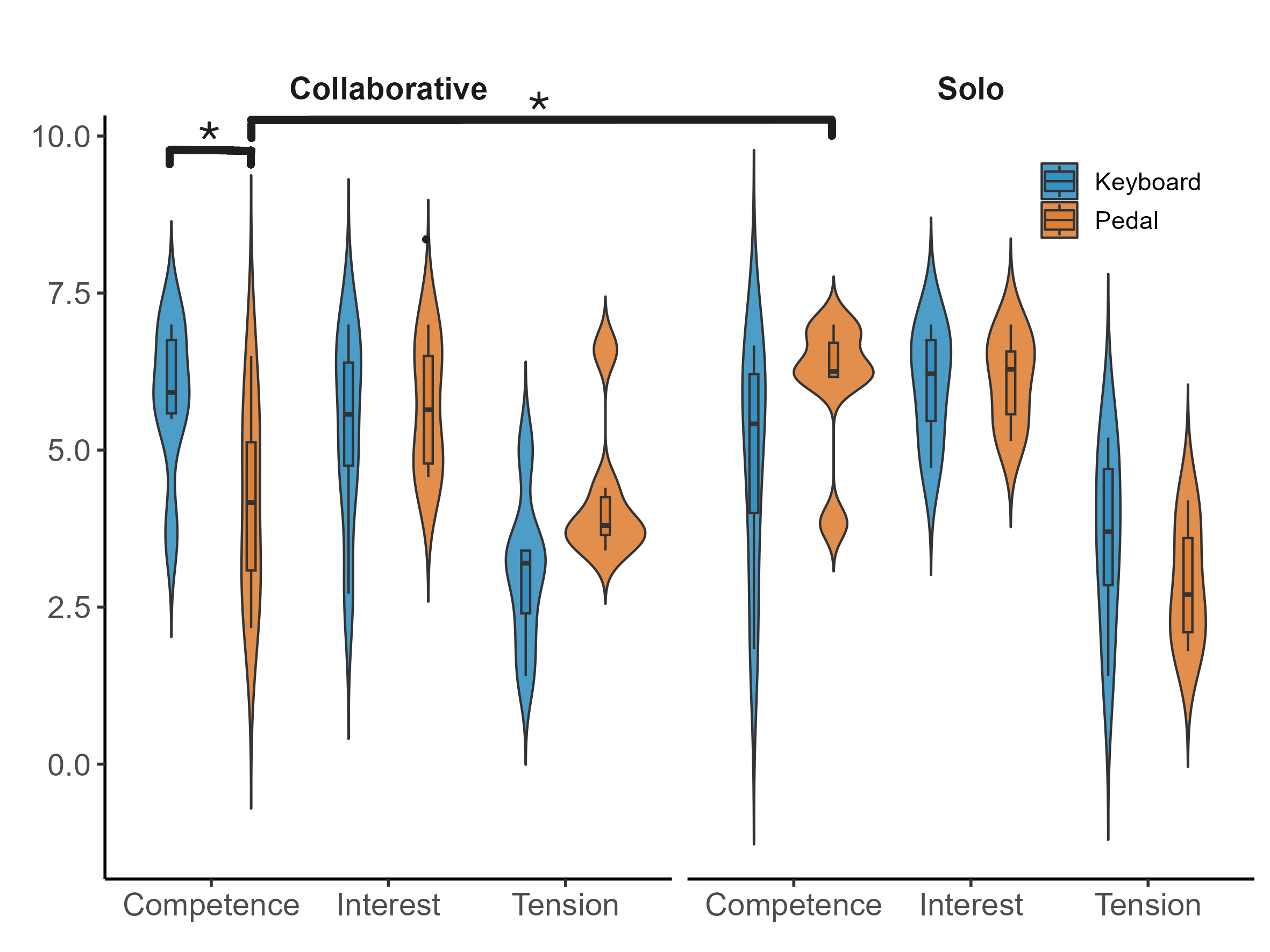}
    \caption{\textbf{PCG Intrinsic Motivation Inventory:} PCG scores trended toward  slightly higher Interest when using the pedal than the keyboard in both game modes. Competence and Tension inversely related to game modes. Statistical significance is shown with an asterisk.}
    \label{fig:imi_cg}%
\end{figure}

As shown in Fig.~\ref{fig:imi_cg}, the median PCG Tension subscale score trended lower when using the pedal ($2.7 \pm 0.98$) than the keyboard ($3.7 \pm 1.44$) in Solo Mode and trended higher when using the pedal ($3.8 \pm 1.19$) than the keyboard ($3.2 \pm 1.22$) in Collaborative Mode. The median PCG Interest subscale score trended higher when using the pedal (Solo: $6.28 \pm 0.73$, Collaborative: $5.64 \pm 1.05$) than the keyboard (Solo: $6.21 \pm 0.90$, Collaborative: $5.57 \pm 1.57$) in both Solo and Collaborative Modes. Finally, the median PCG Competence subscale score trended higher when using the pedal ($6.25 \pm 1.14$) than the keyboard ($5.42 \pm 1.84$) in Solo Mode and was significantly lower when using the pedal ($4.17 \pm 1.62$) than the keyboard ($5.91 \pm 1.23$) in Collaborative Mode ($z = -2.10$, $p = 0.04$). When using the pedal, Competence subscale scores were marginally significantly lower in Collaborative Mode than Solo Mode ($t(5) = -2.57$, $p = 0.05$). See Table~\ref{tab:imi} for all statistical comparisons.

\begin{figure}[t]
    \centering
    \includegraphics[width=1\columnwidth,trim={2pt 0 0 0},clip]{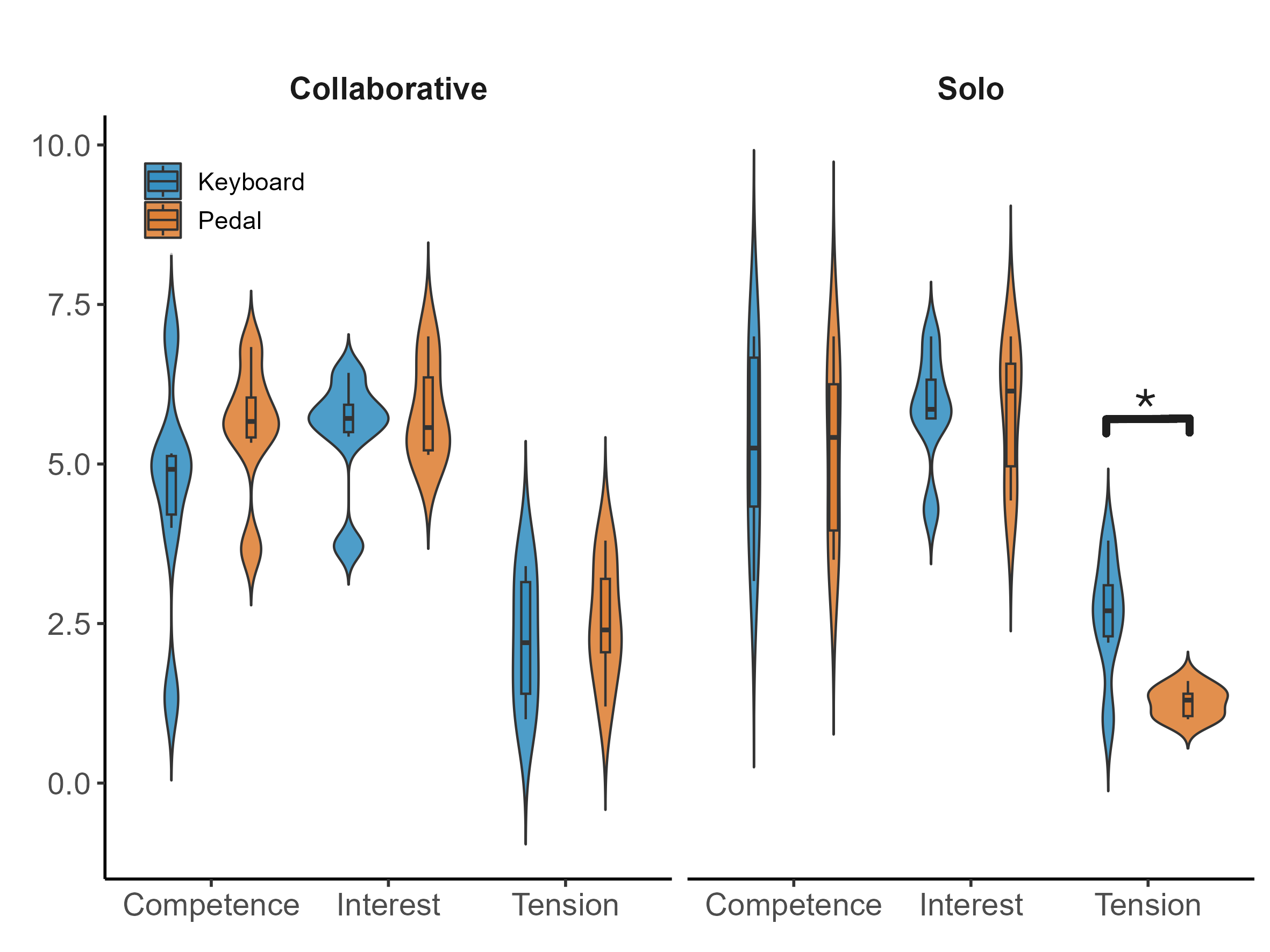}
    \caption{\textbf{PPS Intrinsic Motivation Inventory:} PPS scores trended toward higher Interest when PCGs played with the keyboard Collaboratively than when PCGs played with the pedal in Solo Mode. Competence and Tension inversely related to game modes. Statistical significance is shown with an asterisk.}
    \label{fig:imi_ss}%
\end{figure}

As shown in Fig.~\ref{fig:imi_ss}, the median PPS Tension subscale score was significantly lower when the PCG used the pedal ($1.3 \pm 0.24$) than the keyboard ($2.7 \pm 0.95$) in Solo Mode and trended slightly higher when the PCG used the pedal ($2.4 \pm 0.95$) than the keyboard ($2.2 \pm 1.04$) in Collaborative Mode ($t(5) = 3.37$, $p = 0.02$). The median PPS Interest subscale score trended higher when the PCG used the pedal ($6.14 \pm 1.09$) than the keyboard ($5.86 \pm 0.91$) in Solo Mode and trended lower when the PCG used the pedal ($5.57 \pm 0.78$) than the keyboard ($5.71 \pm 0.94$) in Collaborative Mode. Finally, the median PPS Competence subscale score trended higher when the PCG used the pedal (Solo: $5.42 \pm 1.45$, Collaborative: $5.67 \pm 1.06$) than the keyboard (Solo: $5.25 \pm 1.55$, Collaborative: $4.92 \pm 1.86$) in both Solo and Collaborative Mode. See Table~\ref{tab:imi} for all statistical comparisons.

\subsection{Preference}
Out of the five game play conditions (Collaborative Joystick-Pedal, Collaborative Joystick-Keyboard, Solo Pedal, Solo Keyboard, and Solo Joystick), 3 out of 6 PCGs preferred Solo Keyboard, 2 out of 6 preferred Collaborative Joystick-Pedal, and 1 out of 6 preferred Collaborative Joystick-Keyboard for themselves. PCGs who preferred some form of the Keyboard condition all described prioritizing the easier controllability. However, the PCG who preferred Collaborative Joystick-Keyboard also valued higher scores, saying ``my performance was better which helped us as a whole'' [D3C3]. The two PCGs who preferred the Collaborative Joystick-Pedal condition, both valued the challenging experience saying ``I liked feeling challenged as well! It built my sense of connection by feeling like we were both improving'' [D6C6]. When asked which mode each PCG believed their partner preferred, 3 out of 6 selected Collaborative Joystick-Keyboard, 2 out of 6 selected Collaborative Joystick-Pedal, and 1 out of 6 selected Solo Joystick. The same two PCGs who preferred Joystick-Pedal for themselves also believed that this mode was more fun for their partner. The remaining PCGs who selected some form of Keyboard condition did so because it was easier for their partner or performance was better.

Out of the five game play conditions, 2 out of 6 PPSs preferred Collaborative Joystick-Keyboard, 3 out of 6 preferred Collaborative Joystick-Pedal, and 1 out of 6 preferred Solo Joystick for themselves. Those who preferred the Joystick-Pedal enjoyed either the challenge or the extra activity that this condition required. One of the PPSs said, ``It was a challenge for both participants and allowed for us to play more equally'' [D3S3]. The two PPSs that chose the Collaborative Joystick-Keyboard mode found that it was more fun specifically because the Keyboard was ``easier for my partner'' [D6S6]. Finally, the person who selected Solo Joystick valued the ability to ``not depend on the other person'' [D4S4]. When asked which mode each PPS believed their partner preferred, 1 out of 6 selected Collaborative Joystick-Keyboard, 2 out of 6 selected Collaborative Joystick-Pedal, and 3 out of 6 selected Solo Keyboard. Those who selected some form of the Keyboard condition felt that their partner preferred the easier game play mode because their partner ``likes to win'' [D1S1] or ``isn't a huge fan of games so she definetly[sic] preffered[sic] the simpler interface'' [D2S2]. Those who selected the Collaborative Joystick-Pedal thought that their partners enjoyed the challenge or higher activity level required in this mode.

\section{Discussion}
We conducted a pilot study with 6 healthy dyads to understand the impact of caregiver exercise-based input in a collaborative therapy serious game on performance, positive affect, and motivation. Our results demonstrate encouraging trends that support future work and refinement of this game play to encourage health behavior and positive emotional experience for both members of the dyad.

In terms of performance, PCGs and PPSs performed worse in Collaborative Mode than Solo Mode. Ideally, playing together in Collaborative Mode should have allowed the PPS to improve their performance, thus combating psychological tensions that may arise from Collaborative Mode and incentivizing a collaborative dynamic. Given the added difficulty of Collaborative Mode in this pilot, our participants may have experienced added tensions, which reduced the motivational aspect of Collaborative Mode. In future work with a healthy population, the PPS device should be made more difficult to control in order to increase the motivational aspects of Collaborative Mode and better simulate the experience of a person with stroke.

Despite the added difficulty of the pedal over the keyboard, PCGs' results trended toward more enjoyment from game play with the pedal than the keyboard. In Solo Mode, PCGs' results trended toward higher positive affect scale scores and lower negative affect scale scores when playing with the pedal than the keyboard (Fig.~\ref{fig:panas_cg}). Though not significant, their Interest subscale scores also trended higher when playing with the pedal than the keyboard in both Collaborative and Solo Mode (Fig.~\ref{fig:panas_cg}), which suggests that the pedal input device was more inherently enjoyable. However, there was a reversal in emotional and motivational experience in Collaborative Mode (Fig.~\ref{fig:panas_cg} and Fig.~\ref{fig:imi_cg}). This is in part explained by the 4 out of 6 PCGs who most preferred some form of Keyboard game play condition due to the ease of use. Three of those 4 preferred the Solo Keyboard condition -- where scores were highest -- above all other play conditions, suggesting that these PCGs valued performance most. This was not the case for all PCGs. Two out of 6 PCGs preferred the Joystick-Pedal condition specifically because of the added challenge and interaction of Collaborative Mode. These findings point to the importance of personality type in intrinsic motivation. Additionally, we would expect that in a clinical population of informal caregivers, personal motivation to assist their person with stroke would take precedent above personal performance in a rehabilitation game. The increased, though not significant, Tension subscale scores and significantly lower Competence subscale scores in Collaborative Mode may also be explained by performance anxiety. This is supported by the lower Tension subscale scores with the pedal, a more difficult input device, than the keyboard in Solo Mode, where the PCG did not play with their partner (Fig.~\ref{fig:imi_cg}). Future game development should look to understand the causes of player performance anxiety and include game elements to reduce Tension and increase Competence in Collaborative Mode. By addressing player frustrations, future versions of this game can support improved user experience and avoid potential conflict between players.

Our pilot study findings also showed positive trends in the PPS game experience when their partner used the pedal input device. PPSs' results trended toward higher positive affect scale scores, lower negative affect scale scores (Fig.~\ref{fig:panas_ss}), and higher Competence subscale scores (Fig.~\ref{fig:imi_ss}) when their partner was playing with the pedal than the keyboard in Collaborative Mode. Given that PPSs performed statistically worse in Collaborative than Solo Mode, PPSs valued collaborative play above decreased performance, perhaps because of their increased Competence. PPSs' Interest subscale scores trended higher and Tension subscale scores were lower when their partners played with the pedal than the keyboard in Solo Mode. This may be the result of hearing their partner challenged by the more difficult input device of the pedal, similar to their own difficulties using the joystick with limited mobility. More work is needed to clarify this. In Collaborative Mode, however, PPSs' Interest subscale scores trended lower and Tension subscale scores trended higher when their partners played with the pedal, potentially the result of PPSs noticing the higher tension some of their partners experienced during the Collaborative Joystick-Pedal condition. Evidence of this perceived tension from their partner, can be found in the 4 out of 6 PPSs who thought their partner preferred some form of the Keyboard input device condition most because it was easier to use and their partner performed better. In fact, 2 out of 6 PPSs preferred the Joystick-Keyboard condition most for themselves because it was explicitly easier for their partner. Thus, the perspectives of these PPSs were impacted by their partners' enjoyment. In spite of this, results show that more PPSs preferred the Collaborative Joystick-Pedal condition (3) than PCGs (2), and in general more PPSs preferred a Collaborative Mode, which suggests that PPSs found value in playing together -- even if their scores were lower -- and enjoyed game play where their partners used the exercise-based input of the pedal input device. To this point in Table~\ref{tab:demographics}, two of the PPSs who preferred the Collaborative Joystick-Pedal condition because it was more ``active'' [D0S0] and allowed the dyad to ``... play more equally'' [D3S3] experienced a positive IOS change, indicating an increase in the perception of closeness to their partner.

Limitations of this study include space constraints, which required both members of the dyad to play the game in the same room. While instructed not to look at each other's screens during Solo Mode, verbal interactions between the players may have impacted the results. Given that the PCGs and PPSs performed better in Solo Mode than in Collaborative Mode, future studies should increase the difficulty of the PPS input device and the pedal to increase the motivational aspect of Collaborative mode, where working together benefits the PPS's score. Another limitation of our study was the limited sample size, which reduced our analytical power, and the composition of our participants who were exclusively younger, university students and their partners. Younger people may be more familiar with digital technologies than older adults, which largely constitute the stroke population. Additionally, our sample population may have been driven more by performance than assisting their mobility challenged partner. This would explain why 3 out of 6 PCGs preferred to play alone with the easier input device, the keyboard, where their scores were not brought down by the added challenge of a teammate with limited functionality. Future work should include the clinical population of people with stroke and their informal caregivers, where the motivation to support a loved one, digital literacy, and demographics align with design intent. Additionally, a randomization of both game play mode and individuals who are more intrinsically motivated by individual achievement versus collaboration would clarify the results.

In summary, the results of this pilot study demonstrate encouraging trends on the role of collaborative game play with caregiver exercise-based input towards improved motivation (IMI), emotional experience (PANAS), and closeness (IOS Change). However, a larger study is needed to determine this with more analytical power. Although statistical significance was not reached in our small sample population, we find that both PPSs' and PCGs' game play experience either trended toward higher positive affect scale scores or higher Interest subscale scores when the PCG played with the pedal, exercise-based input, collaboratively. We would expect the benefits of this type of game play to be even more profound in a clinical population where caregivers in a therapeutic game are primarily motivated to support the recovery of the person with stroke. This type of game play could be particularly beneficial for dyads of informal caregiver and people with stroke with strong collectivist motivations \cite{vasquezSocialCulturalFactorsDesign2022} and limited time to care for each other. By improving motivation and supporting both members of the dyad, such a technology, once refined, could support both a higher dosage of therapy for people with stroke and also exercise in caregivers, improving the health of both parties and potentially preventing more strokes within a family unit.

\section*{Acknowledgment}
We thank Y. Meneses for prototyping and J. Muccini, A. Nehrujee, and E. Ivanova for study design advice.

\bibliographystyle{IEEEtran}
\bibliography{references}

\end{document}